\documentclass{article}

\usepackage[utf8]{inputenc}
\usepackage[T1]{fontenc}
\usepackage{amsmath,amssymb}
\usepackage{graphicx}
\usepackage{float}
\usepackage{array}
\usepackage{booktabs}
\usepackage[margin=1in]{geometry}
\usepackage[numbers,sort&compress]{natbib}
\usepackage{url}
\usepackage{wrapfig}
\usepackage[export]{adjustbox}
\usepackage[hidelinks]{hyperref}

\title{Omics Data Discovery Agents: Agent-Supported Retrieval, Reanalysis, and Synthesis of Published Omics Data}

\author{
Alexandre Hutton$^{1}$ \and Jesse G. Meyer$^{1,*}$\\
\\
\small $^{1}$Department of Computational Biomedicine, Cedars-Sinai Medical Center, Los Angeles, CA, USA\\
\small $^{*}$Corresponding author: jesse.meyer@csmc.edu
}

\date{}

\begin{document}

\maketitle


\begin{abstract}
The biomedical literature contains a vast collection of omics studies, yet most published data remain functionally inaccessible for computational reuse. When raw data are deposited in public repositories, essential information for reproducing reported results is dispersed across main text, supplementary files, and code repositories, and in the rarer cases where intermediate data (e.g.\ protein abundance files) are shared, their location is irregular. Here we present an agentic framework for the agent-supported retrieval, reanalysis, and synthesis of published omics data. The system employs large language model (LLM) agents with access to tools for fetching omics studies, extracting article metadata, identifying and downloading published data, executing containerized quantification pipelines, and synthesizing results across studies. Applied at corpus scale, the pipeline catalogued dataset references across thousands of PubMed Central articles; we report these as descriptive system outputs rather than as a validated measure of extraction accuracy. Using model context protocol (MCP) servers to expose containerized analysis tools, the agents retrieved and re-quantified data in five end-to-end reanalyses spanning data-dependent and data-independent proteomics and bulk RNA-seq. All five reanalyses completed, each with documented human guidance and workflow accommodations, and reproduced the authors' deposited abundances with high per-sample correlation (0.85--0.997) and strongly concordant differentially expressed features (fold-change Spearman 0.88--0.91), with no direction reversals among features called differentially expressed in both analyses; residual differences in significant-feature lists were attributable to threshold placement, tool-version, and preprocessing differences rather than to the underlying quantities. We further demonstrate that agents can identify semantically similar studies, judge data compatibility, and synthesize findings across studies, including a random-effects meta-analysis that recovered consistent protein regulation in liver fibrosis. Rather than a validated benchmark of literature-wide performance, this work is a feasibility demonstration together with an auditable, reusable toolset, establishing a foundation for prospective evaluation of automated omics-data reuse.
\end{abstract}
\vspace{1em}

\section*{Author summary}
Every year, tens of thousands of biological studies measure the abundance of thousands of proteins or genes at once, so-called omics experiments, and deposit the resulting data in public archives. In principle these data can be reused to answer new questions or to check whether published findings hold up. In practice, reuse is rare: the information needed to make sense of a dataset, such as which samples are which, how the data were processed, and with which software, is scattered across the article, its supplementary files, and code repositories, and the raw data often must be reprocessed with specialized tools before it can be compared. We built ODDA, a system in which artificial-intelligence language-model agents read the scientific literature, locate the associated data, and reprocess it using the same kinds of containerized tools the original authors used, all through a controlled, auditable interface. Across five studies spanning proteomics and RNA sequencing, ODDA reproduced the authors' published measurements closely and combined results across studies. We share the software so that others can build on, and rigorously test, automated data reuse.

\newpage


\section{Introduction}

The biomedical literature has a growing collection of omics studies, yet the data underlying these publications is impractical for computational reuse. Although raw data are increasingly deposited in public repositories that specialize in distributing them, including PRIDE~\cite{perez2025pride}, MassIVE~\cite{wang2018assembling}, GEO~\cite{edgar2002geo,barrett2012ncbi}, and the Sequence Read Archive~\cite{katz2022sequence}. These resources are a critical component for ensuring that the results reported in scientific articles remain verifiable. However, raw data require domain knowledge to ensure that the quantification tools are set up correctly and that acceptable values are used for their parameters. Despite being the most immediately-useful form of the data, the quantified data are often not placed in public repositories and the barriers to reprocess the data significantly discourages data reuse.

When processed quantitative data are provided, they may appear in different locations and with different levels of completeness. Without a standard way of distributing quantitative data, processing parameters and analytical decisions are similarly scattered across the main article text, supplemental materials, and code repositories. Some journals allow authors to make their data available ``upon request,'' but responses to such requests are frequently incomplete or reveal that the materials are no longer available due to personnel turnover, storage limitations, or similar constraints. A previous study found that data availability declines with article age, with the odds of a dataset still being available falling by 17\% per year after publication~\cite{vines2014availability}.

The cumulative effect is that most published omics studies cannot be computationally reused without substantial manual effort. In practice, this limits reproducibility, cross-study synthesis, downstream discovery, and interdisciplinary collaboration. The scale of the problem precludes manual solutions: in November 2025 alone, more than 1,500 articles matching the query ``proteomics OR mass spectrometry'' were available from PubMed Central (PMC). A similar volume of articles are available for genomics, transcriptomics, and metabolomics.

Current large-scale automated curation efforts primarily focus on linking existing author-provided metadata, which is often limited to a small set of keywords. Commercial tools that use AI assistants also exist, but are limited to summarizing literature or generating data-analysis code, rather than identifying datasets and making them accessible via natural language. Platforms such as ScienceMachine~\cite{sciencemachine} or Kosmos~\cite{kosmos} perform custom, on-demand analysis of individual studies rather than indexing and re-quantifying the large corpus of data held in public omics repositories. To our knowledge, no existing system combines these capabilities at corpus scale: autonomously identifying relevant omics studies, recovering their underlying data and metadata, executing appropriate analyses on demand, and returning auditable, study-level results. Recent advances in large language models (LLMs) and their application to knowledge-graph construction~\cite{pan2024unifying} suggest new possibilities for automated literature processing. Emerging work has demonstrated the potential of AI agents for biomedical data analysis~\cite{huang2025biomni}, though these approaches have not yet been combined with comprehensive literature mining to utilize the value of existing biomedical omics publications.

Here, we present an agentic framework that addresses these limitations by automating the identification, extraction, and linking of omics research products from full-text articles and supplemental material. Unlike static metadata-harvesting approaches used by existing repositories, we extract information from the unstructured publication text and link the article to its datasets, sample annotations, processing methods, and code repositories. The structured representation enables searches beyond keyword matching, including semantic queries that better capture article content. A central feature is the deployment of AI agents that can locate relevant studies, identify quantitative datasets, fetch and process raw data in a manner similar to the original publication, and integrate results across multiple studies. The extracted information allow agents to reason over the full publication record, including the main text, supplementary files, repository metadata, and available code. When processed data is available, we demonstrate that agents are able to identify and locate it whether it is in supplemental materials, standard repositories, or project-specific sites. When only raw data are available, agents infer the processing context from available documentation and invoke standard quantification tools with parameters derived from the publication.

We demonstrate the feasibility of this approach through three core capabilities: (1) corpus-scale metadata extraction from articles available via PMC and insertion into a relational database, reported here as descriptive system outputs rather than validated accuracy; (2) agent-guided raw-data reanalysis using containerized tools exposed through model context protocol (MCP) servers; and (3) cross-study reasoning that identifies compatible datasets and synthesizes findings across publications. Our results demonstrate that, with documented human guidance, LLM agents can transform unstructured omics publications into auditable, re-executable research objects, and that agent-supported reanalysis can recover reproducible results across studies, including findings not reported in the original articles. Rather than establishing literature-wide performance, this work provides a feasibility demonstration and an open, auditable toolset that lay a foundation for prospective, larger-scale evaluation of automated omics-data reuse.

\section{Methods}

\subsection{System Architecture Overview}

\begin{figure}[t]
    \centering
    \includegraphics[width=\linewidth]{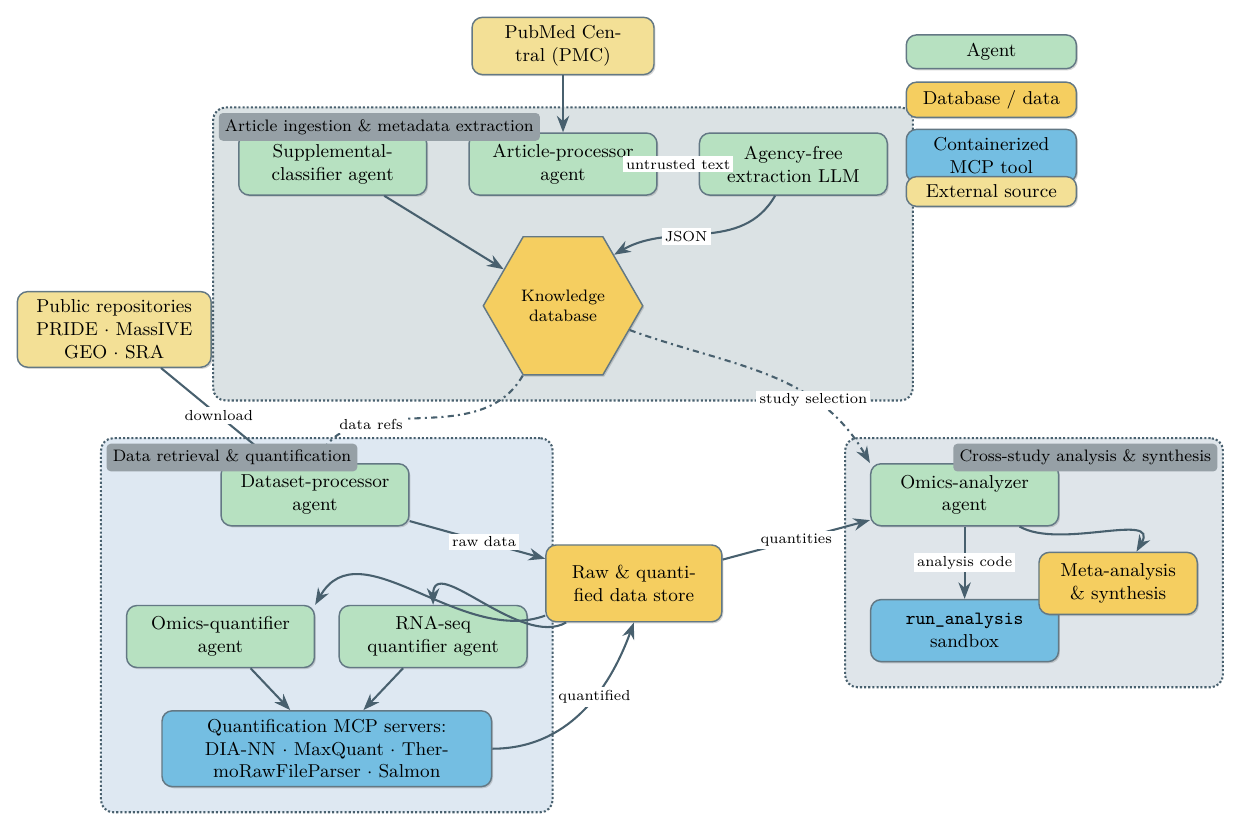}
    \caption{Schematic of the ODDA agent system. Published literature (PubMed Central) is ingested by the article-processor and supplemental-classifier agents; untrusted article text is routed to a separate, agency-free extraction LLM that returns JSON, which is written to a provenance-stamped knowledge database (reader/actor separation). Guided by that database, the dataset-processor agent retrieves raw and processed data from public repositories (PRIDE, MassIVE, GEO, SRA); the omics-quantifier and RNA-seq quantifier agents re-quantify it through containerized MCP servers (DIA-NN, MaxQuant, ThermoRawFileParser, Salmon) into a raw-and-quantified data store. The omics-analyzer agent then performs cross-study analysis and random-effects meta-analysis, executing any synthesized code only in the least-privilege \texttt{run\_analysis} sandbox. Node colors denote component type (legend).}
    \label{fig:schematic}
\end{figure}

Our agentic curation and reanalysis system consists of three components, shown in Figure~\ref{fig:schematic}: (1) an article ingestion and metadata extraction pipeline, (2) a data retrieval and quantification system, (3) a cross-study reasoning system that identifies compatible datasets and synthesizes findings across multiple publications.

\subsection{Article Ingestion and Metadata Extraction}
\subsubsection{Data Sources}

We retrieved full-text articles from the PubMed Central Open Access (PMC OA) subset using the NCBI E-utilities API. We identified articles returned from the query, "proteomics" for November 2025 - January 2026 in addition to "liver fibrosis proteomics" for 2025, for a total of 4210 articles, with the bulk of the articles coming from the broader "proteomics" search.

\subsubsection{LLM-Based Information Extraction}

Metadata extraction is done in two steps: gathering existing structured metadata such as title, journal, keywords, etc. that are readily available via PMC XML fields, and gathering unstructured metadata from the full text and supplemental. The full article text was processed using a separate large language model to extract additional keywords and identify the locations of raw data, analysis code, and quantitative data. The extraction model used in this work was OpenAI's GPT-5 accessed through Azure AI Foundry; because the provider and model are user-configurable (Methods, ``model layer''), the exact resolved model identifier, provider, and timestamp are recorded in the provenance layer for every extraction, so that the specific model version and the date of each retrieval are self-documenting rather than fixed in prose. The article-retrieval dates reported here (November 2025--January 2026) refer to the PMC query window used to assemble the corpus. The purpose of the separate LLM is to limit the possible effects of prompt injections; the external LLM is instructed to return a JSON-formatted response, and the content is processed by functions that do not expose the contents to the agent. The separation can be achieved via programmatic access to models, which is currently readily available from the major providers (e.g. Anthropic's Client SDKs, Azure's AI Foundry) or executed locally via Ollama~\cite{noauthor_ollama_nodate}.
Agents were instructed to provide evidence citations for all extracted values and to explicitly abstain when information could not be reliably determined. Text embeddings were generated using OpenAI's text-embedding-3-small model and stored in a database for use with semantic similarity search.

\subsection{Corpus-scale extraction outputs}

The extraction pipeline was applied to the full article corpus to catalogue references to raw and processed omics data, and we report the resulting article counts as descriptive outputs of the system. We did not benchmark extraction precision or recall against an independently annotated gold standard in this work. The corpus counts may therefore include both false positives (spurious or misclassified references) and omissions, and are not intended as estimates of prevalence or of the system's extraction accuracy. Establishing extraction accuracy against an independent, adjudicated benchmark is left to prospective evaluation.

\subsection{Agent-Guided Raw Data Reanalysis}

\subsubsection{Model Context Protocol (MCP) Server Design}

Analytical tools were made available to agents via function calls provided by MCP servers. This design prevents agents from developing custom tools for each article to ensure reproducibility through containerized execution. MCP servers expose the following method categories:

\textbf{Data retrieval methods:} Transfer data from public data repositories (PRIDE, MassIVE, GEO), inspect file structure and formats.

\textbf{Quantification pipeline selection and configuration:} Using descriptions from the article and available metadata, identify appropriate quantification pipelines (e.g. DIA-NN, MaxQuant) and ensure that assumptions made by pipelines are consistent with data acquisition parameters. Once identified, a configuration file for use in later quantification steps is generated.

\textbf{Execution methods:} Launch containerized analysis pipelines (Apptainer) using information from the previously-generated configuration files. Estimate resource usage and enforce limits. Analysis code derived from untrusted article text is executed only through a least-privilege sandbox (see Discussion and the repository threat model), never on the host or via a general shell.

\subsubsection{Containerized Analysis Tools}

For proteomics data, the MCP server exposes two containerized quantification tools for the main data acquisition types: DIA-NN~\cite{demichev2020diann} for data-independent acquisition (DIA) and MaxQuant~\cite{tyanova2016maxquant} for data-dependent acquisition (DDA). These workflows produce protein-level abundance matrices mapped to standardized UniProt~\cite{uniprot2025} gene identifiers. For transcriptomics data, the server exposes a containerized Salmon~\cite{patro2017salmon} tool that builds a decoy-aware transcriptome index and quantifies raw RNA-seq reads to transcript- and gene-level abundances (TPM and read counts), demonstrating that the same MCP architecture generalizes across omics modalities. All quantification tools are distributed as Apptainer definition files with reproducible build scripts; because several tools carry redistribution-restricted licenses, we provide the definitions and versioned build recipes rather than the prebuilt images.

\subsubsection{Agent Workflow for Quantification}

When an article is identified for quantification, database tools first verify whether raw data (e.g. mass spectrometry) has been identified. If so, an agent operates the following workflow:

\begin{enumerate}
    \item \textbf{Parameter recovery:} First check whether the processing parameters were already extracted from the article's full text during ingestion (and recorded in the database), and whether the raw data or supplementary material includes a deposited parameter or configuration file; only otherwise does the agent parse the article text, supplementary materials, and code repositories to identify analysis parameters including enzyme specificity, missed cleavages, mass tolerances, variable modifications, and software versions.
    \item \textbf{Data retrieval:} Use the data retrieval MCP tools to download raw data files from identified repositories.
    \item \textbf{Tool selection:} Based on the data acquisition type and article description, select the appropriate containerized analysis tool.
    \item \textbf{Configuration:} Generate a configuration file specifying all identified parameters. Parameters not explicitly stated in the article are flagged as defaults.
    \item \textbf{Execution:} Invoke the containerized pipeline and monitor for completion or errors.
    \item \textbf{Validation:} Compare output dimensions and summary statistics against article-reported values when available.
\end{enumerate}

\subsubsection{Evaluation of Quantification}

We evaluated reanalysis performance using articles where both raw and processed data were publicly available. We define a \emph{successful completion} as an end-to-end run comprising raw-data retrieval, parameter recovery from the article, tool selection, containerized execution, sample matching to the deposited design, and comparison with the deposited quantities. We distinguish successful completion from fully autonomous execution by recording, for each study, the human guidance and workflow accommodations required: software-version substitutions, prompts to match the article's tool version, instrument-format conversions, disabled options, and normalization decisions. Where a deposited matrix had unlabeled sample columns (GSE272283), samples were mapped to columns using the study's GEO/SRA metadata rather than by expression correlation. All attempted reanalyses are reported; none were excluded on the basis of outcome. Primary metrics included correlation between agent-generated and published abundances and overlap in differentially expressed features when statistical analysis was performed. To attribute the sources of any discrepancy rather than reporting overlap alone, we decomposed differences in differentially expressed features into (i) features identified by one analysis but not the other, (ii) fold-change magnitude differences relative to the calling threshold, and (iii) genuine reversals in the direction of regulation. We additionally report the Spearman and Pearson concordance of fold changes between the original and reanalyzed data, and the sensitivity of set overlap to the significance and fold-change thresholds. Protein-group tables were aggregated to gene level before comparison (gene-symbol key; protein groups mapping to more than one gene were excluded rather than duplicated; unambiguous groups mapping to the same gene were summed on the linear intensity scale). Differential-expression significance was assessed with an imputed pipeline (median normalization, down-shifted-normal imputation, Welch $t$-test, Benjamini--Hochberg), whereas fold-change concordance and direction agreement were computed from a separate non-imputed matrix, using observed log2 fold changes only for genes with sufficient observations in both conditions of both datasets. Where the original analysis used a software version that could not be redistributed as a container, we matched the article's preprocessing parameters (recovered from deposited configuration and log files) using the nearest available containerized version and report the version difference explicitly.

\subsubsection{Cross-study meta-analysis}
For quantitative synthesis across studies, per-protein effect sizes (log2 fold change, fibrotic versus control) and their standard errors were obtained in one of two ways. For studies with replicate-level quantities (Cheng and Jirouskova), the effect size was the difference of group means of gene-level log2 abundances, and its standard error was computed directly from the replicate variances, $\mathrm{SE}=\sqrt{s^2_{\mathrm{case}}/n_{\mathrm{case}}+s^2_{\mathrm{ctrl}}/n_{\mathrm{ctrl}}}$. For the Devos summary-statistic datasets, which report only a fold change and a two-sided uncorrected p-value per dataset, the standard error was inferred from the p-value using a standard-normal approximation, $\mathrm{SE}=|\widehat{\log_2 FC}|/\Phi^{-1}(1-p/2)$. Because Devos et al. assembled several distinct, independently-sourced fibrosis datasets, each dataset was treated as an independent study-level estimate. Effect sizes were combined with fixed-effect and DerSimonian--Laird random-effects meta-analysis, reporting pooled estimates with 95\% confidence intervals, a two-sided p-value, and between-study heterogeneity (Cochran's $Q$, $I^2$, and $\tau^2$). This procedure is implemented as a reusable MCP tool so that any assembled set of comparable studies can be synthesized in the same way.


\subsection{Cross-Study Reasoning}

\subsubsection{Study Compatibility Assessment}

For cross-study analyses, an agent was instructed to obtain 3 articles examining similar problems, as determined by the cosine similarity between the text embeddings of their abstracts. When available, the agent was instructed to use already-quantified data. Otherwise, it was to obtain the public raw data and quantify it as described in the corresponding article.


\section{Results}
\subsection{Corpus-scale article analysis}
Applying the ingestion pipeline to the assembled corpus, ODDA retrieved 4,210 articles, of which 2,442 had accessible full text. Within the full-text set, the system identified references to published raw data in 1,264 articles, references to processed quantitative data in 710 articles, and references to both in 464 articles. These are system-generated counts produced directly by the extraction pipeline: they describe the volume of dataset references that ODDA surfaced across the corpus and were not validated against an independent, manually curated gold standard. They should therefore be read as descriptive outputs of the system, not as estimates of prevalence, precision, recall, or data-recovery rates.

\subsection{End-to-end reanalysis of published omics data}

We next assessed whether agents could reproduce published omics results by retrieving raw data and re-quantifying it with containerized tools. We report five end-to-end reanalyses spanning two proteomics acquisition types and one transcriptomics modality: data-dependent acquisition (DDA) proteomics (MaxQuant), data-independent acquisition (DIA) proteomics (DIA-NN, two studies), and bulk RNA-seq (Salmon, two studies). These five studies were the complete set we attempted; none were excluded on the basis of failure or poor agreement, so the results below are worked demonstrations of feasibility rather than a randomly sampled benchmark.

\begin{figure}[H]
    \centering
    \includegraphics[width=0.95\linewidth]{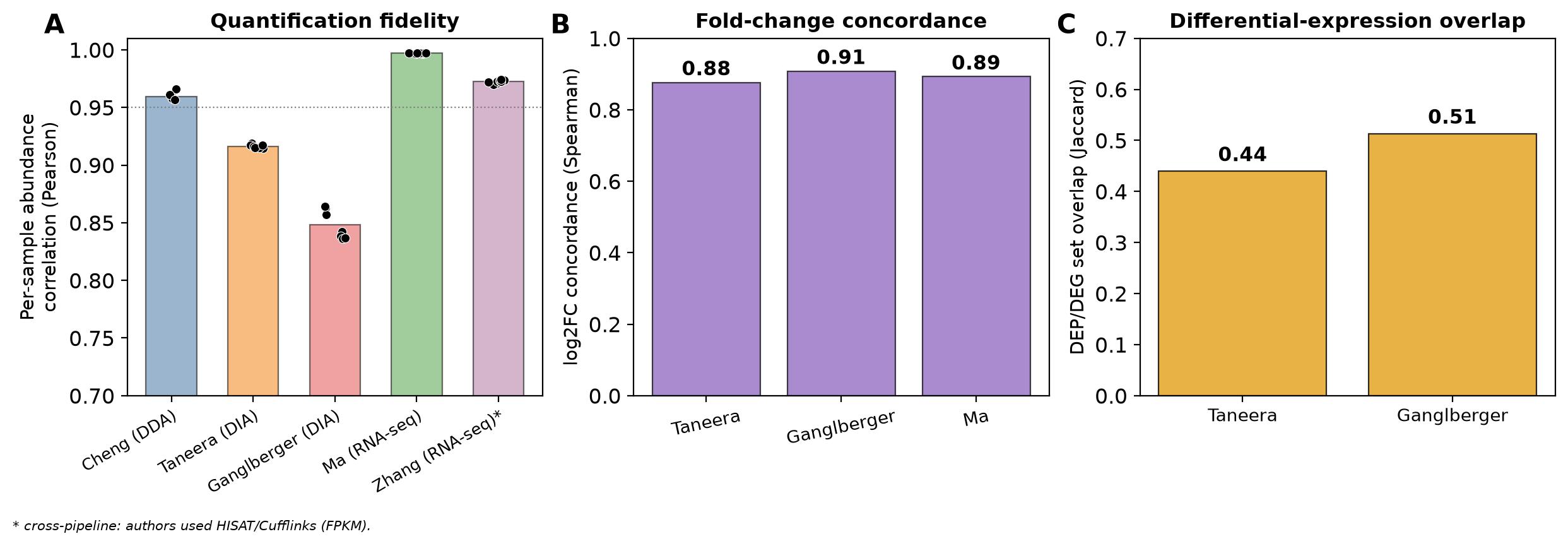}
    \caption{Reproduction fidelity across the five studies. (A) Per-sample abundance correlation between the ODDA re-quantification and the authors' deposited values (bars, mean; points, individual samples). (B) Concordance of case/control log2 fold changes (Spearman). (C) Overlap of significant DEP/DEG sets (Jaccard). Fidelity is high across proteomics and transcriptomics despite tool-version and pipeline differences; these are worked demonstrations, not a literature-wide estimate.}
    \label{fig:fidelity}
\end{figure}

We define a \emph{successful completion} as an end-to-end run in which the agent retrieved the raw data, recovered the processing parameters from the article, selected an appropriate tool, executed the containerized pipeline, matched samples to the deposited design, and compared its output against the authors' deposited quantities. Successful completion in this sense is distinct from fully autonomous execution: every reanalysis required some human guidance or workflow accommodation: a substituted software version, a prompt to match the article's tool version, an instrument-format conversion, a disabled option, or a normalization decision, which we record per study in Table~\ref{tab:fidelity} and in the detailed narratives below. Under this definition all five reanalyses completed, recovering the authors' deposited abundances with high per-sample correlation (Pearson 0.85--0.997) and, where differential expression was performed, strongly concordant fold changes (Spearman 0.88--0.91), with no reversals in direction among features called differentially expressed in both analyses (Figure~\ref{fig:fidelity}).

\begin{table}[H]
\centering
\footnotesize
\setlength{\tabcolsep}{3pt}
\caption{Central validation table: the five end-to-end reanalyses attempted, none excluded. ``Source'' is the repository from which raw data were retrieved; ``Versions'' gives the article's tool version and the containerized version used by ODDA. ``Human intervention'' records the guidance or workflow accommodation required, distinguishing successful completion from fully autonomous execution. Abund.\ $r$ is the mean per-sample abundance correlation (Pearson, log scale); FC $\rho$ is the Spearman correlation of case/control log2 fold changes; Overlap is the Jaccard index of significant-feature sets (adjusted $p<0.05$, $|\log_2\mathrm{FC}|>1$). Dashes indicate a metric not applicable to that study's comparison.}
\label{tab:fidelity}
\begin{tabular}{@{}>{\raggedright\arraybackslash}p{1.55cm}>{\raggedright\arraybackslash}p{1.4cm}>{\raggedright\arraybackslash}p{1.0cm}>{\raggedright\arraybackslash}p{1.6cm}>{\raggedright\arraybackslash}p{3.0cm}cccc@{}}
\hline
\textbf{Study} & \textbf{Modality / Tool} & \textbf{Source} & \textbf{Versions (orig.\ $\to$ ODDA)} & \textbf{Human intervention} & \textbf{Compl.} & \textbf{Abund.\ $r$} & \textbf{FC $\rho$} & \textbf{Overlap} \\
\hline
Cheng et al. & DDA / MaxQuant & iProX & 2.1.4.0 $\to$ 2.8.1.0 & Version substitution; match-between-runs and second-peptide search disabled (FAIMS limit of buildable version) & \checkmark & 0.96$^{a}$ & -- & -- \\
Taneera et al. & DIA / DIA-NN & PRIDE & 1.8.1 $\to$ 2.6.1 & Version substitution; prompted to match article version; instrument RAW converted to mzML & \checkmark & 0.92 & 0.88 & 0.44 \\
Ganglberger et al. & DIA / DIA-NN & PRIDE & 1.9.2 $\to$ 2.6.1 & Version substitution; instrument RAW converted to mzML & \checkmark & 0.85 & 0.91 & 0.51 \\
Ma et al. & RNA-seq / Salmon & SRA &1.1.0 $\to$ 2.3.3 & Version substitution; reused decoy-aware index & \checkmark & 0.997 & 0.89 & --$^{b}$ \\
Zhang et al. & RNA-seq / Salmon & SRA &HISAT/\allowbreak Cufflinks $\to$ 2.3.3 & Cross-tool substitution; unit matching (TPM vs.\ FPKM); genotype/column mapping from GEO/SRA metadata; one run re-downloaded & \checkmark & 0.97$^{c}$ & -- & -- \\
\hline
\end{tabular}
\\[2pt]{\footnotesize $^{a}$Identification Jaccard 0.90 for the shared-protein set. $^{b}$No genes reached the standardized DEG threshold (common CPM/Welch pipeline, adjusted $p<0.05$, $|\log_2\mathrm{FC}|>1$) at the profiled 1-week stage in either re-analysis; this is a like-for-like re-analysis, not a reproduction of the authors' original DESeq2 calls. $^{c}$Cross-tool correlation of ODDA gene-level TPM against the authors' deposited FPKM (like-with-like length-normalized units), a lower bound on same-tool reproduction.}
\end{table}

The detailed reanalyses follow.

\subsubsection{Cheng et al.\ (DDA proteomics)}
\begin{figure}[H]
    \centering
    \includegraphics[width=0.5\paperwidth]{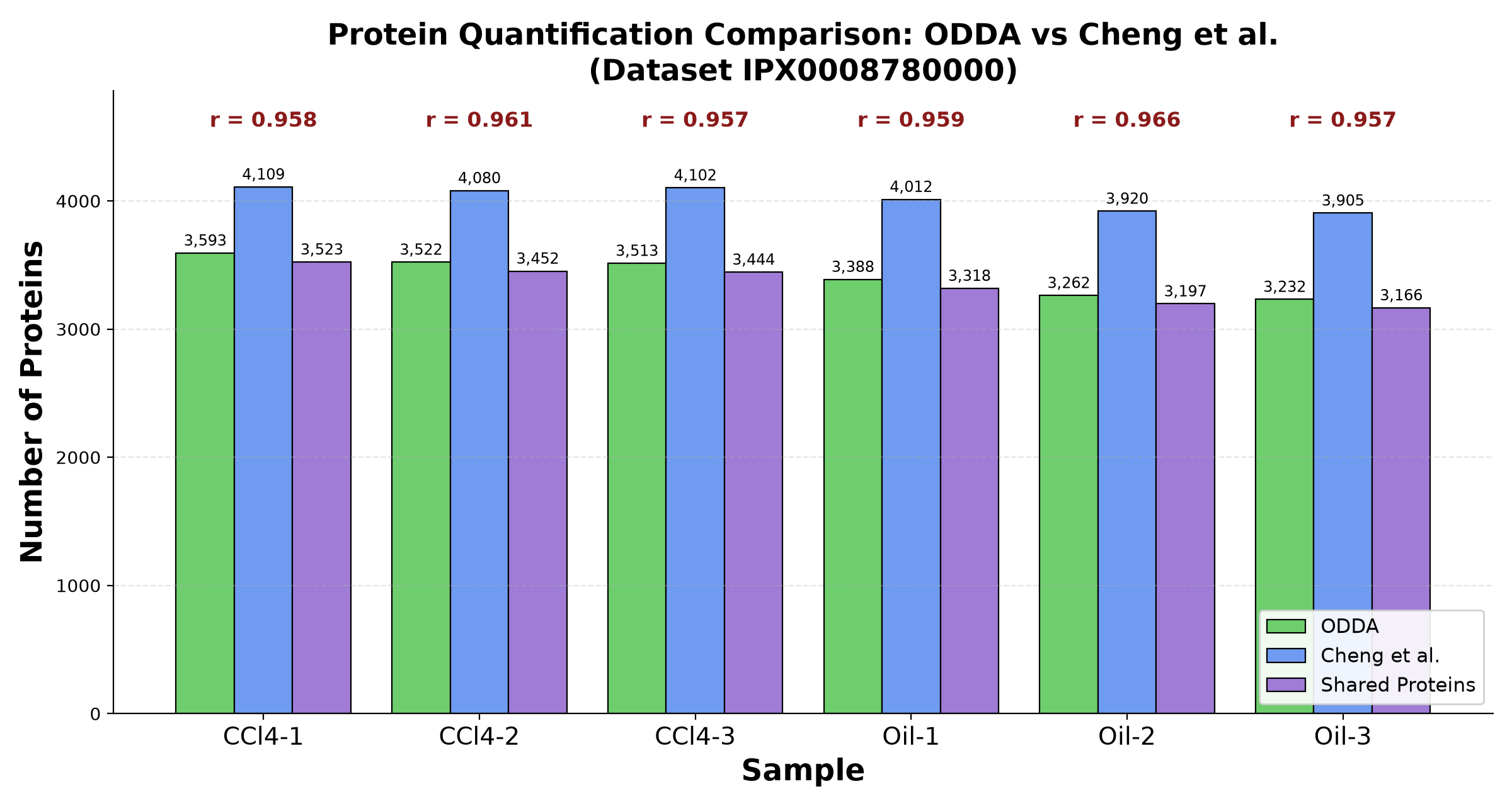}
    \caption{Comparison of protein identifications reported by Cheng et al. and those obtained by ODDA using the article content to inform parameter selection. Per-sample counts of proteins with valid intensity for ODDA (MaxQuant v2.8.1.0, dual-pass disabled owing to a FAIMS limitation of that release) and the published data, together with the number of shared proteins and the per-sample Pearson correlation of log-transformed intensities for the shared set. ODDA identifies slightly fewer proteins per sample while the overlapping quantities remain strongly correlated (r$>$0.95).}
    \label{fig:protein_comparison}
\end{figure}

The article by Cheng et al.~\cite{cheng2025proteomics} provides its quantitative data as well as its raw data, making it well-suited for comparing results. The raw files were fetched by the agent via ProteomeXchange~\cite{Vizcaino2014-un}, which linked to the files hosted on IProX~\cite{Ma2019-ug}. The files were quantified with the containerized MaxQuant tool using a parameter file generated by the agent from values obtained during article ingestion, matching the authors' deposited settings (trypsin with up to two missed cleavages; fixed carbamidomethyl (C); variable oxidation (M) and protein N-terminal acetylation; 20~ppm tolerances; 1\% FDR at the PSM, protein, and site levels). We note an important reproducibility constraint: the article was processed with MaxQuant v2.1.4.0 (recovered from the parameter file deposited alongside the raw data), but the only version we were able to rebuild as a redistributable container was v2.8.1.0, and this release cannot perform match-between-runs or second-peptide search on FAIMS-acquired data (both raise runtime errors on FAIMS runs). Disabling these ``dual-pass'' options (which normally transfer identifications across runs and recover additional low-abundance peptides) directly accounts for the modestly lower number of identifications reported by ODDA. After quantification, both the article and ODDA datasets were filtered to remove reverse hits, contaminants, and proteins identified only by site; proteins were matched by UniProt Majority Protein ID and by gene name; and the number of proteins with valid intensity values was counted for each of the 6 samples (CCl4-1/2/3, Oil-1/2/3). Intensities were log-transformed and Pearson correlation was calculated per-sample and pooled across all samples. ODDA identified 4,179 protein groups versus 4,406 in the published data (identification Jaccard index 0.90); although the published analysis quantified somewhat more proteins per sample (consistent with the disabled match-between-runs), the quantities of the overlapping proteins were highly correlated (per-sample Pearson 0.957--0.966, pooled 0.960; Figure~\ref{fig:protein_comparison}).

\subsubsection{Taneera et al.\ (DIA proteomics)}
The source article~\cite{taneera2026ppp1r1a} reported several of the parameters required for processing, including the quantification method used. Key parameters included trypsin enzyme specificity (stated in the article), a data-independent acquisition (DIA) mode, DIA-NN as the quantification software, and two experimental conditions (control and PPP1R1A knockdown), which were identified from the raw-data filenames.

Notably, the agent differentiated between the two experimental conditions (control vs. knockdown) based on filename patterns in the raw data without explicit instruction, demonstrating autonomous interpretation of experimental design when the correspondence between experimental conditions and files is not explicitly stated.

The article reported 9{,}006 quantified proteins; re-quantifying the raw DIA data with a containerized DIA-NN (v2.6.1) and aggregating protein groups to the gene level (Methods) yielded 8{,}175 gene-level features. We then investigated differential expression between the control and PPP1R1A-knockdown conditions; the agent recovered the article's statistical approach (Benjamini--Hochberg correction) from its methods section and reproduced the preprocessing before testing. At the standard thresholds (adjusted $p<0.05$ and $|\log_2\mathrm{FC}|>1$), the differentially expressed protein (DEP) sets from the deposited data and our re-analysis overlapped only partially (Jaccard 0.44; 56 significant in the deposited data, 39 in the re-analysis, 29 shared), lower than perfect concordance, so we investigated the source of the discrepancy.

\begin{figure}[H]
    \centering
    \includegraphics[width=0.9\linewidth]{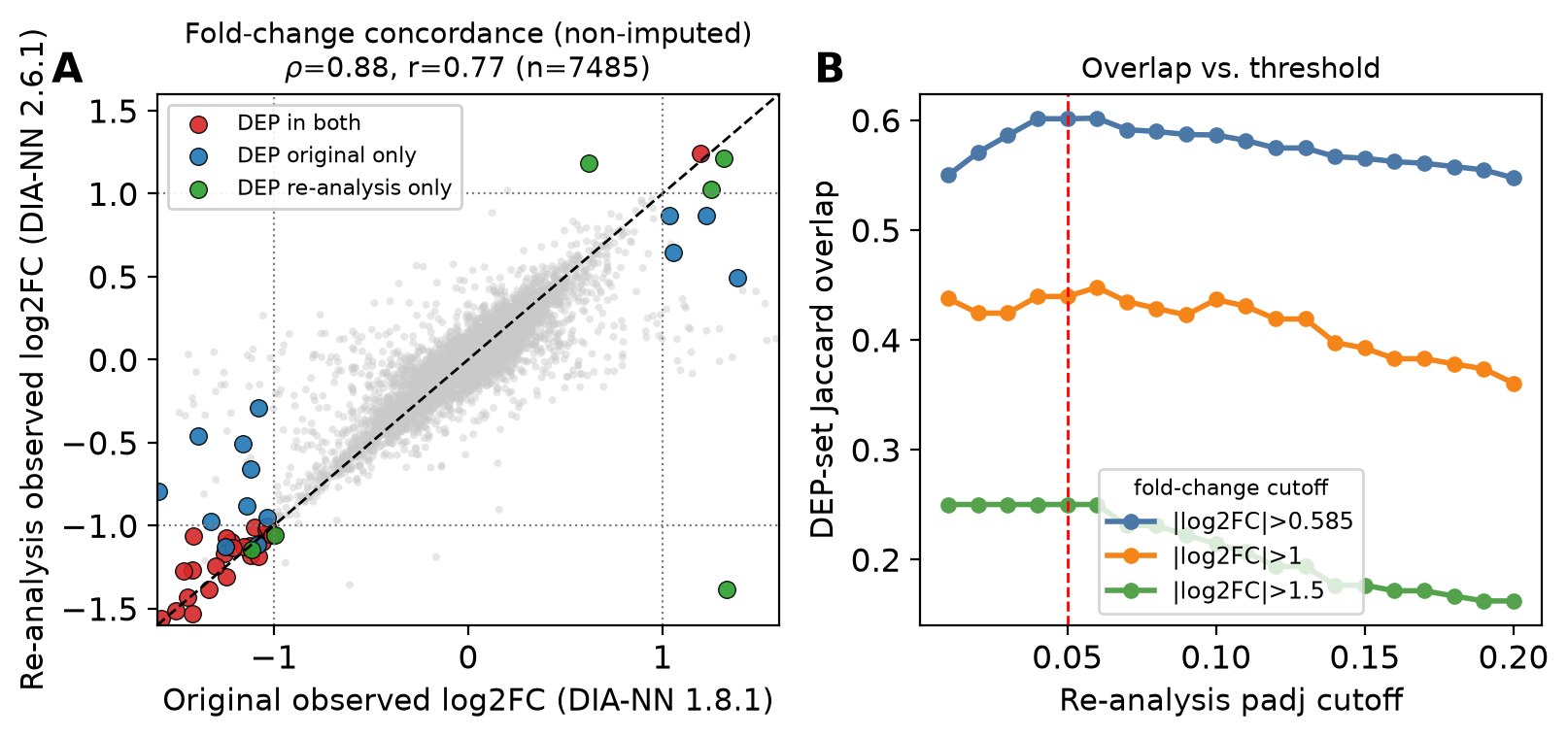}
    \caption{Reproduction of the Taneera et al. differential-expression analysis (siPPP knockdown vs. siNC control). (A) Per-gene log2 fold changes from the original deposited quantification (DIA-NN 1.8.1) versus the ODDA re-quantification (DIA-NN 2.6.1); differentially expressed genes are colored by whether they are called in both analyses, only the original, or only the re-analysis. Observed (non-imputed) fold changes are strongly concordant (Spearman $\rho=0.88$ across the 7{,}485 genes that met the observation rule) and no gene significant in both analyses reverses its direction of regulation. (B) DEP-set overlap (Jaccard) as a function of the re-analysis significance threshold for three fold-change cutoffs, showing that the incomplete overlap is governed by threshold placement rather than quantitative disagreement.}
    \label{fig:taneera_dep}
\end{figure}

For this decomposition we re-quantified the raw DIA files with the latest available containerized DIA-NN (v2.6.1), matching the authors' preprocessing recovered from their deposited DIA-NN log (rat reference proteome; trypsin with one missed cleavage; fixed carbamidomethylation; N-terminal methionine excision; 1\% precursor and protein-group FDR; match-between-runs), aggregated protein groups to genes, and applied two complementary pipelines to both matrices (Methods): an imputed pipeline for differential-expression significance calls (median normalization, down-shifted-normal imputation, Welch $t$-test, Benjamini--Hochberg correction) and a separate non-imputed pipeline for fold-change concordance (observed log2 fold changes computed only for genes with sufficient observations in both conditions of both datasets). The residual difference was largely attributable to threshold placement rather than to disagreement in the underlying quantities: the observed fold changes were strongly concordant across the 7{,}485 genes that met the observation rule (of 7{,}665 quantified in both matrices; Spearman $\rho = 0.88$, Pearson $0.77$), no protein called differentially expressed in \emph{both} analyses reversed its direction of regulation (one robustly-changed protein, EARS2, reversed sign but was significant in only one analysis), and of the 27 originally-significant DEPs not re-called, 18 failed the two-fold fold-change cutoff (with the same direction of change) and 9 failed the adjusted-p cutoff. Relaxing the fold-change threshold from two-fold to 1.5-fold raised the Jaccard index to approximately 0.60 (Figure~\ref{fig:taneera_dep}). The incomplete overlap therefore reflects borderline proteins near the significance and fold-change boundaries, an expected consequence of the DIA-NN version difference (the authors used v1.8.1) combined with our emulation of the original R-based imputation and testing in Python, rather than a failure to recover the biological signal.

\subsubsection{Ganglberger et al.\ (DIA proteomics)}
As a second, independent DIA proteomics test (PXD063923; mouse models of congenital stationary night blindness~\cite{ganglberger2025csnb2}), we re-quantified the raw data with DIA-NN and compared against the authors' deposited protein matrix using the same gene-level, two-pipeline procedure. Per-sample gene-level abundances correlated at Pearson 0.85, and for the disease-versus-wildtype contrast the observed (non-imputed) log2 fold changes were highly concordant across the 3{,}822 genes that met the observation rule (of 4{,}561 quantified in both matrices; Spearman 0.91, Pearson 0.92). The differential-expression signal was far larger than in the Taneera knockdown (1{,}625 significant genes in the deposited data and 2{,}086 in the re-analysis, overlapping at Jaccard 0.51); of the 367 originally-significant genes not re-called, 260 failed the adjusted-p cutoff and 107 the fold-change cutoff. As in the Taneera comparison, no gene called differentially expressed in \emph{both} analyses reversed its direction of regulation; among the wider set of single-analysis calls, 18 genes changed sign, almost all at near-zero magnitude (median per-reversal minimum $|\log_2\mathrm{FC}|$ of 0.12, i.e. the median of the smaller of the two estimates). The residual set-overlap difference is therefore largely concentrated at the calling threshold rather than reflecting quantitative disagreement, although 275 union DEPs lacked sufficient non-imputed observations to assess and two robustly-changed genes (PGK2, PWWP3A) reversed sign between the single-analysis calls.

\subsubsection{Ma et al.\ (RNA-seq)}
\begin{figure}[H]
    \centering
    \includegraphics[width=0.9\linewidth]{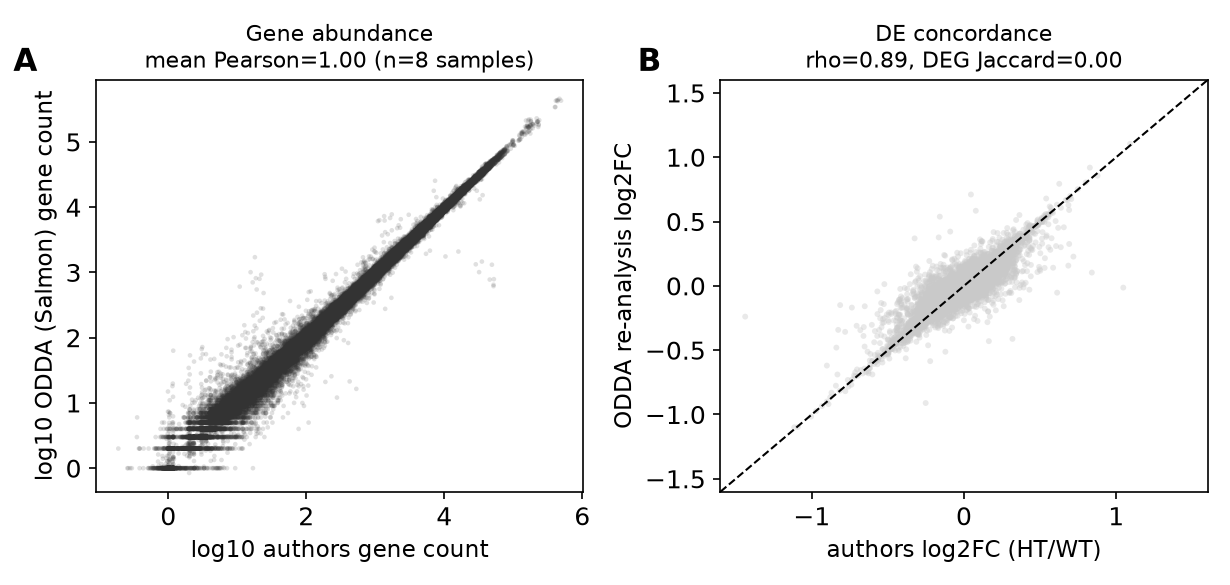}
    \caption{Transcriptomics reproduction (GSE292102). (A) Gene-level abundance for the ODDA Salmon re-quantification versus the authors' deposited counts (pooled over 8 samples); the near-identity relationship gives a mean per-sample Pearson of 0.997. (B) Concordance of genotype log2 fold changes.}
    \label{fig:rnaseq}
\end{figure}

As a demonstration that the same MCP architecture ports beyond proteomics, we reanalyzed two bulk RNA-seq studies end-to-end from raw FASTQ: reads were retrieved from the SRA, quantified with a containerized Salmon decoy-aware index built from the GENCODE mouse transcriptome, and summarized to gene level, then compared against the authors' deposited count matrices. These two studies are illustrative of portability rather than a systematic transcriptomics benchmark. The Cyfip2 developmental study (GSE292102~\cite{ma2025cyfip2}) deposited a 48-sample series spanning three ages (1, 7, and 14 weeks), two brain regions (cortex and hippocampus), and two genotypes (Cyfip2 heterozygote and wild type) in quadruplicate; our reproduction used the eight-sample 1-week cortex subset (heterozygote versus wild type; runs SRR32732030--037, GSM8849242--249; Methods), the earliest developmental cortical stage on which the study centers. For this subset, the ODDA gene-level quantification reproduced the authors' deposited (DESeq2-normalized) count matrix almost exactly (mean per-sample Pearson 0.997, Spearman 0.999 across more than 18{,}000 genes; Figure~\ref{fig:rnaseq}), and the genotype log2 fold changes were concordant (Spearman 0.89). Applying a single standardized pipeline (CPM normalization, Welch $t$-test, adjusted $p<0.05$, $|\log_2\mathrm{FC}|>1$) to both matrices, no gene reached this threshold in either re-analysis; this is a like-for-like re-analysis of the two matrices rather than a reproduction of the authors' original DESeq2 test (adjusted $p<0.05$). These values apply to the specified subset, not the full GSE292102 series.

\subsubsection{Zhang et al.\ (RNA-seq)}
The second RNA-seq study (GSE272283, a CCl4 liver-fibrosis Mfap2-knockout model~\cite{zhang2025mfap2}) reinforces the paper's liver-fibrosis theme. Its eight samples are the 8-week CCl4 liver-fibrosis experiment (Mfap2 knockout versus wild type, four each; not the separate 4-week-recovery series). The deposited count-matrix columns are labeled only as Sample~1--8, but the study's GEO/SRA metadata assign them unambiguously (Samples 1--4 wild type, Samples 5--8 knockout), so runs were mapped to columns directly from that metadata (Methods). The authors quantified with a different toolchain (HISAT alignment and Cufflinks, reporting FPKM). Comparing our Salmon gene-level TPM against the deposited FPKM (matching length-normalized units), ODDA recovered the abundances with a mean per-sample correlation of 0.97 across the eight samples. This cross-tool concordance is a conservative lower bound on same-tool reproduction and shows that the transcriptomics reproduction is robust even when the original and re-analysis pipelines differ in both aligner and quantifier. Comparing our raw counts to their FPKM (mismatched units) reduced this correlation to 0.87, underscoring the importance of matching the reported normalization when reproducing deposited results.

\subsection{Cross-Study Compatibility Assessment}

We evaluated the system's ability to identify compatible studies for data integration. When presented with two studies examining proteomics of liver pathology (one one liver cancer, the other on fatty liver disease), the agent correctly determined that the datasets were not directly comparable and declined to perform integration.

\begin{figure}[H]
    \centering
    \includegraphics[width=0.5\linewidth]{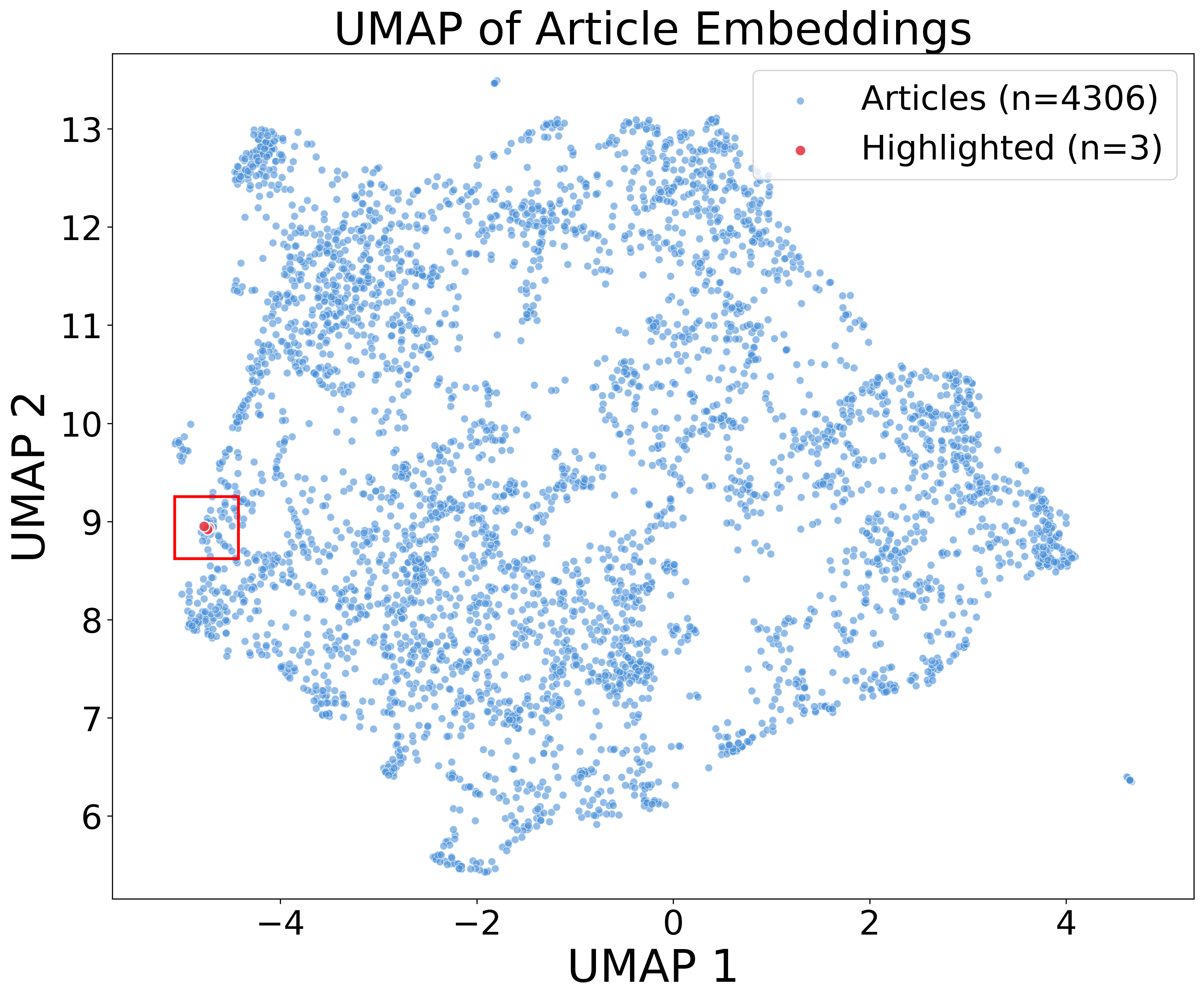}
    \caption{Distribution of articles in a UMAP. Text embeddings were computed from article abstracts and projected with UMAP. The projection includes all articles in the database with computed embeddings ($n=4{,}306$); this exceeds the 4{,}210-article retrieval corpus reported above because the database also contains articles ingested later for other work, and the semantic-similarity search operated over this full embedding set. The three articles identified as semantically similar are highlighted in red.}
    \label{fig:article_umap}
\end{figure}

When tasked to find 3 similar articles using the minimum pairwise distance measured by the cosine similarity of the abstract embeddings, the agent identified three studies: Cheng et al.~\cite{cheng2025proteomics}, Jirouskova et al.~\cite{jirouskova2025dynamics}, and Devos et al.~\cite{Devos2026-qx}. All three papers investigated proteomics in liver fibrosis. Cheng et al. examined proteomics and phosphoproteomics to show kinase dysregulation in mice. Jirouskova et al. reported on the role of clusterin, an extracellular chaperone protein. Lastly, Devos et al. reported on differentially expressed proteins in both heart and liver fibrosis in human patients, identifying 18 DEPs shared across the organs. We visualize their proximity using a UMAP of all embedded articles in the database (Figure~\ref{fig:article_umap}).

The agent obtained protein quantities for the raw data shared by Cheng et al. by using MaxQuant as described in the article; differential expression was obtained by comparing across the case/control files. Case vs. control was automatically determined by considering the experimental procedure in the article and the filenames (\_CCL4\_ vs. \_OIL\_). The protein quantities for Jirouskova et al. were supplied in the public repository, and those were used for comparison. The 18 DEPs reported by Devos et al. are not listed in the main article, but are found in a nested archive within the supplemental material that was automatically downloaded during article ingestion.
When tasked with determining whether the articles demonstrated agreement, the agent obtained the list of 18 proteins from Devos et al. and compared these with the differential expression reported in the data from Cheng et al. and Jirouskova et al., matched at the gene-symbol level, and found that 10/18 proteins showed the same direction of regulation (upregulation) across species. Of those 10, 8 had an evaluable case/control effect in all three study groups (CLU, LUM, AMBP, COL6A2, MFAP4, MYH10, PRELP, and COL14A1). The other two were directionally concordant but not evaluable in all three: TGFBI was quantified in every group but was absent from the Cheng control samples, so no Cheng fold change could be computed, and FBLN5 was detected in only one of the mouse-based experiments. Although the direction of regulation agreed across studies, the degree of regulation was considerably different for most proteins.
We note that other than the upregulation of CLU reported as one of the main findings in Jirouskova et al., the \textbf{remaining proteins are not reported} in the main text of any of the three articles. To determine that the three studies demonstrate reproducible differential expression, it was necessary to obtain full article, determine processing parameters, quantify one of the datasets, identify proteins of interest, and compare high-level differential expression for the proteins.
Literature review confirmed that the directionally-concordant proteins have previously reported roles in fibrosis: CLU~\cite{cells8111442,wang_clusterin_2026}, TGFBI~\cite{krzistetzko_association_2023,wu_tgfbi_2026}, AMBP~\cite{hakuno_hepatokine_2018}, MYH10~\cite{liu_distinct_2011}, PRELP~\cite{yamauchi_elevated_2024}, lumican (LUM)~\cite{krishnan_lumican_2012}, collagen~VI (COL6A2)~\cite{williams_collagenvi_2022}, and MFAP4~\cite{saekmose_mfap4_2015}; COL14A1 has been reported as differentially deposited during the fibrosis-to-carcinoma transition~\cite{lai_extracellular_2011}.

\subsubsection{Quantitative cross-study meta-analysis}
\begin{figure}[H]
    \centering
    \includegraphics[width=0.98\linewidth]{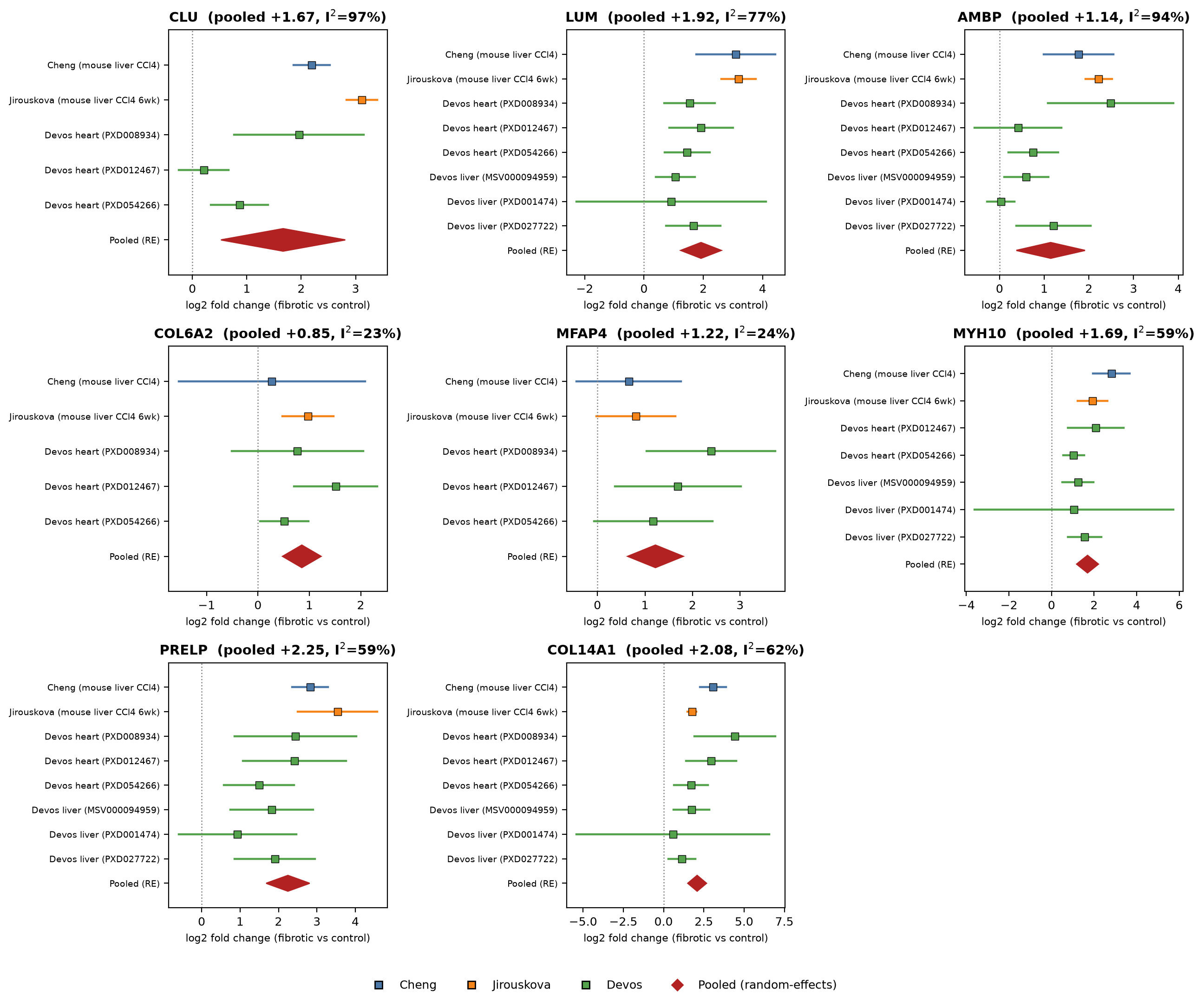}
    \caption{Random-effects meta-analysis of the eight fibrosis proteins with an evaluable effect in all three study groups. Each panel shows the available per-study/per-dataset log2 fold-change estimates (fibrotic vs.\ control; squares with 95\% confidence intervals); the number of contributing estimates $k$ ranges from 5 to 8, since some Devos datasets did not quantify a given protein. Estimates from Cheng et al.\ (mouse liver CCl4) and Jirouskova et al.\ (mouse liver CCl4, 6-week) use replicate-derived standard errors; the independent human datasets assembled by Devos et al.\ (heart fibrosis, HF vs.\ non-HF: PXD008934, PXD012467, PXD054266; early-versus-severe liver fibrosis, F3/F4 vs.\ F0: MSV000094959, PXD001474, PXD027722) use standard errors inferred from reported p-values. The pooled random-effects estimate is shown as a red diamond. All eight proteins are significantly up-regulated in fibrosis; the between-study heterogeneity ($I^2$) reflects magnitude differences across models and species despite a consistent direction of regulation.}
    \label{fig:forest}
\end{figure}

To move beyond a qualitative statement of directional agreement, we treated the cross-study comparison as a formal random-effects meta-analysis. For each consistently-regulated protein we assembled effect sizes (log2 fold change, fibrotic versus control) with their standard errors from each study: Cheng et al.~\cite{cheng2025proteomics} (mouse CCl4 liver, from our re-quantification) and Jirouskova et al.~\cite{jirouskova2025dynamics} (mouse CCl4 liver, 6-week timepoint, from the deposited MaxQuant quantities), both with replicate-derived standard errors, together with the distinct, independently-sourced human fibrosis datasets that Devos et al.~\cite{Devos2026-qx} assembled (three heart-fibrosis datasets, HF versus non-HF; and three early-versus-severe liver-fibrosis datasets, F3/F4 versus F0), each treated as an independent study-level estimate with a standard error inferred from its reported p-value (see Methods). Effect sizes were combined using the DerSimonian--Laird random-effects estimator, taking each study's available estimate (the number of contributing datasets $k$ ranged from 5 to 8 per protein). All eight proteins evaluable in every study group were significantly up-regulated in fibrosis, with pooled log2 fold changes of CLU $+1.67$ (95\% CI $0.53$--$2.81$), LUM $+1.92$ ($1.24$--$2.60$), AMBP $+1.14$ ($0.38$--$1.91$), COL6A2 $+0.85$ ($0.48$--$1.23$), MFAP4 $+1.22$ ($0.64$--$1.80$), MYH10 $+1.69$ ($1.18$--$2.20$), PRELP $+2.25$ ($1.68$--$2.81$), and COL14A1 $+2.08$ ($1.51$--$2.66$); every pooled 95\% confidence interval excluded zero ($p<5\times10^{-3}$; Figure~\ref{fig:forest}). The between-study heterogeneity ($I^2$ = 23--97\%) formally quantifies the earlier observation that the \emph{direction} of regulation is consistent while its \emph{magnitude} differs across models, organs, and species. This synthesis is exposed as a reusable meta-analysis MCP tool, so the same quantitative combination can be applied to any set of studies the system assembles. We present it as a demonstration of automated cross-study synthesis on a curated three-study set rather than as a broad meta-analytic benchmark.

\subsection{Summary of demonstrated capabilities}

Across the tasks we evaluated, the agentic system was able to:

\begin{enumerate}
    \item \textbf{Recover processing parameters from article text} sufficient to reproduce published analyses, with the required human guidance and workflow accommodations documented per study.
    \item \textbf{Reproduce deposited quantities with high fidelity} in five worked reanalyses: strongly correlated abundances for shared features (per-sample Pearson 0.85--0.997; identification Jaccard 0.90 for Cheng et al.) and, where differential expression was performed, concordant fold changes (Spearman 0.88--0.91) with no direction reversals among features significant in both analyses; incomplete significant-feature overlap was attributable to threshold placement, tool-version, and preprocessing differences rather than to lost signal.
    \item \textbf{Assess study compatibility:} identification of comparable studies and appropriate refusal to integrate incompatible datasets.
    \item \textbf{Synthesize findings across studies:} cross-species directional agreement for liver-fibrosis proteins, formalized by a random-effects meta-analysis in which all eight proteins evaluable across the three study groups showed significant pooled up-regulation (pooled log2FC 0.85--2.25, all $p<0.005$), presented as a demonstration of cross-study synthesis rather than a broad benchmark.
\end{enumerate}

\section{Discussion}

\subsection{Significance for Omics Data Reuse}
The availability of publicly accessible datasets provides an opportunity to verify the results of new experiments using previously published data. While a single article is generally narrow in scope, the associated data may contain information relevant to other questions. Finding relevant data is challenging, especially when only raw data is available, and quantification would otherwise require substantial manual intervention. Agentic systems provide a new opportunity to convert unstructured descriptions of scientific articles into searchable datasets by automating many labor-intensive aspects of data processing.

\subsection{Technical Innovations}
The main contribution of this manuscript is to demonstrate that datasets contain rich, untapped information that cannot be captured by examining only the article's main text. Extracting a more-complete metadata requires examining supplemental materials and uploaded datasets. By providing tools accessible via MCP servers for transferring, examining, and processing datasets, agentic systems can be used to obtain that information reliably and transparently. Since most tasks do not require novel analyses (e.g., interacting with APIs), the use of MCP servers reduces the risk that agents execute code with silent errors.

\subsection{Reproduction fidelity in the studies examined}
Across the five re-quantification studies we attempted, spanning DDA and DIA proteomics and bulk RNA-seq, the agentic re-analysis recovered the authors' deposited quantities with high fidelity (per-sample abundance correlation 0.85--0.997) and produced strongly concordant differential-expression fold changes (Spearman 0.88--0.91), with no direction reversals among features called differentially expressed in both analyses. Where the categorical lists of differentially expressed features overlapped only partially (Jaccard 0.44--0.51), the difference was in every case dominated by borderline features near the significance and fold-change thresholds rather than by disagreement in the underlying measurements; relaxing the fold-change cutoff substantially increased the overlap. A recurring and important theme is that exact reproduction is bounded by the availability of the original software version: several tools cannot be redistributed or rebuilt at the exact version an article used, and successive releases change identification, false-discovery-rate, and quantification models in ways that shift borderline features across thresholds. Rather than obscure this, the provenance-stamped design lets ODDA measure and attribute the gap, separating tool-version effects and threshold placement from genuine analytical disagreement, which we suggest is an appropriate practice for automated reuse. The same containerized MCP interface extended without modification from proteomics (MaxQuant, DIA-NN) to transcriptomics (Salmon), demonstrating that the architecture ports across these omics modalities in the cases examined; we treat this as a demonstration of portability rather than evidence of universal generalization.

\subsection{Limitations}
Several limitations bound the scope of the claims we make. First, extraction accuracy was not independently benchmarked in this work: the corpus-scale article counts are descriptive system outputs and may contain both false positives and omissions, so they do not establish the system's precision, recall, or data-recovery rate. Second, the five reanalyses were selected worked examples rather than a randomly sampled set; five studies cannot estimate literature-wide success, and the reproduction fidelity we report should not be read as a general performance floor. Third, none of the reanalyses was fully autonomous: each required documented human guidance and workflow accommodations (software-version substitutions, prompts to match tool versions, instrument-format conversions, disabled options, and normalization decisions), so ``successful completion'' as we define it is distinct from unattended operation. Establishing extraction accuracy against an independently annotated benchmark and evaluating reanalysis success on a larger, prospectively sampled set of studies are the natural next steps that this framework and its released tooling are intended to support.

The high-level comparison across studies (e.g. differential expression) requires the execution of experiment-specific code to consider the findings of interest; implementing static MCP-based analyses would require adherence to standardized data structures and would only be applicable to a subset of experiments. Extending agent functionality to multi-study comparisons inherently requires a flexible treatment of experimental data, which in turn introduces the possibility of idiosyncratic analyses.

MCP servers are not ultimate safeguards: they simply provide an interface for pre-defined tools. Malicious or insecure implementations of an MCP server still presents a risk. We note that this is not limited to the system proposed in this article but is a known risk with LLMs using MCP servers.

As with any use of LLMs, prompt injection is a possible attack vector, and it is not one that can be fully eliminated: an LLM that must read attacker-influenced text can in principle be steered by it. During testing, obvious attacks such as local files with fake PMCIDs were detected when data verification failed (e.g., no article was found for the fake PMCID), and similarly for a real PMCID carrying metadata that did not match the article contents. However, adding a prompt into a real article was successful and went unnoticed: for example, a sentence instructing the system to extract specific keywords successfully introduced the requested keywords into the database. It is unlikely that such an injection would survive in the main text of a published article, as it would be visible during peer review or after publication; the agent system, however, also examines supplementary files, which reviewers are less likely to scrutinize.

Rather than attempt to detect hostile phrasing, which an informed attacker can rewrite to evade, we mitigate this risk architecturally by separating reading from acting. Agents that can call MCP tools and potentially execute code do not read potentially-hostile text directly; instead they invoke tools that send the text to a separate LLM without agency, tools, or memory, and that model can only return a JSON document that is parsed and stored by deterministic code without the agent ever seeing the raw text. An injected instruction can therefore still corrupt the classification of the record being processed (e.g. spurious keywords or a mislabeled file), but its effect is bounded to that single record and cannot reach other records, the tools, or the host. We regard bounded, per-record metadata poisoning of this kind as a residual risk to be surfaced by human review rather than one that any in-band text filter can reliably prevent.

The greatest risk is in cross-study synthesis, where the agent inspects the article text to determine how to reproduce the analysis and, in doing so, emits specialized code; a prompt injection at that stage could steer that code toward malicious behavior. We note that the secure way to execute potentially-malicious code is not to do so, and we advise close inspection of the produced analyses for reasons of both security and scientific understanding. Where such code must be run, it is confined to a least-privilege sandbox (detailed in the repository threat model): an Apptainer container with no new privileges, a read-only root filesystem and a single scoped scratch mount, network egress disabled by default (which neutralizes the exfiltration and download-and-run vectors), bind mounts limited to the specific dataset under analysis with no access to credentials, other datasets, or the database, and enforced CPU, memory, wall-clock, and output-size limits. Because this containment is applied by the execution tool rather than requested of the agent, it holds even if the agent is manipulated. We are candid about the boundary of this guarantee: an agent asked only by prompt to run its code in the sandbox could ignore the instruction, so the sandbox must be the enforced execution path rather than a convention the agent is trusted to follow. Fully addressing prompt injection at scale would require sandboxing every agent, not only the code-synthesizing one, so that each can reach only the LLM endpoint and a mediated data-fetch interface; we implement the highest-risk case here and regard comprehensive per-agent sandboxing as the principled next step.

\subsection{Summary}
We presented ODDA, an agentic framework that converts unstructured omics publications into provenance-stamped, re-executable research objects. Alongside corpus-scale metadata extraction, we showed in five worked reanalyses that agents operating containerized MCP tools can re-quantify published data across proteomics and transcriptomics with high fidelity to the authors' deposited results, attribute residual differences to tool-version, threshold, and preprocessing effects rather than analytical error, and synthesize findings across related studies, each with the required human guidance documented. By making the location, retrieval, and reanalysis of published omics data auditable and reproducible, this work provides a feasibility demonstration and an open toolset that lay a foundation for prospective, larger-scale evaluation of automated omics-data reuse.

\section*{Funding}
This work was supported by the National Institute of General Medical Sciences (NIGMS) of the National Institutes of Health under award number R35GM142502 to J.G.M. The funders had no role in study design, data collection and analysis, decision to publish, or preparation of the manuscript. The funding was used to support the salary of both authors. 

\section*{Author Contributions}
Conceptualization, data curation, formal analysis, investigation, methodology, software, validation, visualization, writing - reviewing and editing: A.H. and J.G.M.; writing - original draft: A.H.; project administration, resources, supervision, funding acquisition: J.G.M.

\section*{Data Availability}
All code for the agentic curation system, MCP servers, and evaluation scripts are available on \href{https://www.github.com/xomicsdatascience/odda}{GitHub}. The database schema and containerized tool definitions (Apptainer definition files) are released under GPL v3.0 license. Due to license restrictions on the quantification tools, we are unable to provide the containers directly, but we have included instructions for users on how to obtain the relevant files while respecting license terms.

\section*{Competing Interests}
The authors declare no competing interests.

\section*{Use of AI Tools}
Claude (Anthropic) was used to draft and revise the text of this manuscript and to generate the underlying ODDA agents and software system. All figures and analyses presented here were produced by Claude and the ODDA system. GPT was used to review all code relative to text and figures, and the authors reviewed and verified all text and figures, and the authors take full responsibility for the content.


\end{document}